\PassOptionsToPackage{unicode}{hyperref}
\PassOptionsToPackage{hyphens}{url}
\documentclass[
  table,
  10pt,
  a4paper]{article}
\usepackage{amsmath,amssymb}
\usepackage{lmodern}
\usepackage{ifxetex,ifluatex}
\ifnum 0\ifxetex 1\fi\ifluatex 1\fi=0 % if pdftex
  \usepackage[T1]{fontenc}
  \usepackage[utf8]{inputenc}
  \usepackage{textcomp} % provide euro and other symbols
\else % if luatex or xetex
  \usepackage{unicode-math}
  \defaultfontfeatures{Scale=MatchLowercase}
  \defaultfontfeatures[\rmfamily]{Ligatures=TeX,Scale=1}
\fi
\IfFileExists{upquote.sty}{\usepackage{upquote}}{}
\IfFileExists{microtype.sty}{% use microtype if available
  \usepackage[]{microtype}
  \UseMicrotypeSet[protrusion]{basicmath} % disable protrusion for tt fonts
}{}
\makeatletter
\@ifundefined{KOMAClassName}{% if non-KOMA class
  \IfFileExists{parskip.sty}{%
    \usepackage{parskip}
  }{% else
    \setlength{\parindent}{0pt}
    \setlength{\parskip}{6pt plus 2pt minus 1pt}}
}{% if KOMA class
  \KOMAoptions{parskip=half}}
\makeatother
\usepackage{xcolor}
\IfFileExists{xurl.sty}{\usepackage{xurl}}{} % add URL line breaks if available
\IfFileExists{bookmark.sty}{\usepackage{bookmark}}{\usepackage{hyperref}}
\hypersetup{
  pdftitle={Viash: from scripts to pipelines},
  pdfauthor={Robrecht Cannoodt1,2,3,; Hendrik Cannoodt1; Eric Van de Kerckhove1; Andy Boschmans1; Dries De Maeyer4; Toni Verbeiren1,},
  hidelinks,
  pdfcreator={LaTeX via pandoc}}
\usepackage[margin=1in]{geometry}
\usepackage{graphicx}
\makeatletter
\def\maxwidth{\ifdim\Gin@nat@width>\linewidth\linewidth\else\Gin@nat@width\fi}
\def\maxheight{\ifdim\Gin@nat@height>\textheight\textheight\else\Gin@nat@height\fi}
\makeatother
\setkeys{Gin}{width=\maxwidth,height=\maxheight,keepaspectratio}
\makeatletter
\def\fps@figure{htbp}
\makeatother
\usepackage{tcolorbox}
\usepackage{colortbl}
\usepackage{booktabs}
\usepackage{tabularx}
\usepackage{pifont}
\usepackage{float}
\usepackage{caption}
\ifluatex
  \usepackage{selnolig}  % disable illegal ligatures
\fi
\usepackage[sorting=none,url=false]{biblatex}
\title{Viash: from scripts to pipelines}
\author{Robrecht Cannoodt\textsuperscript{1,2,3,*} \and Hendrik
Cannoodt\textsuperscript{1} \and Eric Van de
Kerckhove\textsuperscript{1} \and Andy
Boschmans\textsuperscript{1} \and Dries De
Maeyer\textsuperscript{4} \and Toni Verbeiren\textsuperscript{1,*}}
\date{21 October 2021}

\begin{document}
\maketitle

\textsuperscript{1} Data Intuitive, Lebbeke, Belgium\\
\textsuperscript{2} Data Mining and Modelling for Biomedicine group, VIB
Center for Inflammation Research, Ghent, Belgium\\
\textsuperscript{3} Department of Applied Mathematics, Computer Science,
and Statistics, Ghent University, Ghent, Belgium\\
\textsuperscript{4} Discovery Sciences, Janssen Research \& Development,
Pharmaceutical Companies of Johnson \& Johnson, Beerse, Belgium

\textsuperscript{*} Correspondence:
\href{mailto:robrecht@data-intuitive.com}{Robrecht Cannoodt
\textless{}\href{mailto:robrecht@data-intuitive.com}{\nolinkurl{robrecht@data-intuitive.com}}\textgreater{}},
\href{mailto:toni@data-intuitive.com}{Toni Verbeiren
\textless{}\href{mailto:toni@data-intuitive.com}{\nolinkurl{toni@data-intuitive.com}}\textgreater{}}

\hypertarget{abstract}{%
\section{Abstract}\label{abstract}}

Most bioinformatics pipelines consist of software components that are
tightly coupled to the logic of the pipeline itself. This limits
reusability of the individual components in the pipeline or introduces
maintenance overhead when they need to be reimplemented in multiple
pipelines. We introduce Viash, a tool for speeding up development of
robust pipelines through ``code-first'' prototyping, separation of
concerns and code generation of modular pipeline components. By
decoupling the component functionality from the pipeline logic,
component functionality becomes fully pipeline-agnostic, and conversely
the resulting pipelines are agnostic towards specific component
requirements. This separation of concerns improves reusability of
components and facilitates multidisciplinar and pan-organisational
collaborations. It has been applied in a variety of projects, from
proof-of-concept pipelines to supporting an international data science
competition.

Viash is available as an open-source project at
\href{https://github.com/viash-io/viash}{github.com/viash-io/viash} and
documentation is available at \href{https://viash.io}{viash.io}.

\hypertarget{introduction}{%
\section{Introduction}\label{introduction}}

Recent developments in high-throughput RNA sequencing and imaging
technologies allow present-day biologists to observe single-cell
characteristics in ever more detail
\autocite{luecken_currentbestpractices_2019}. As the dataset size and
the complexity of bioinformatics workflows increases, so does the need
for scalable and reproducible data science. In single cell biology,
recent efforts to standardise some of the most common single-cell
analyses {[}nf-core, VSN pipelines{]} tackle these challenges by using a
combination of Nextflow
\autocite{ditommaso_nextflowenablesreproducible_2017}, containerisation
(e.g.~Docker, Singularity) and horizontal scaling in cloud computing
(e.g.~AWS, Azure). Further optimization is possible through the use of
hardware acceleration \autocite{avantika_lal_2021_4638196}.

Since research projects are increasingly more complex and
interdisciplinary, researchers from different fields and backgrounds are
required to join forces. This implies that not all project contributors
can be experts in computer science. The chosen framework for such
projects therefore needs to have a low barrier to entry in order for
contributors to be able to participate. One common pitfall which greatly
increases the barrier to entry is tightly coupling a pipeline and the
components it consists of. Major drawbacks include lower transparency of
the overall workflow, limited reusability of pipeline components,
increased complexity, debugging time increase and a greater amount of
time spent refactoring and maintaining boilerplate code. Non-expert
developers in particular will experience more arduous debugging sessions
as they need to treat the pipeline as a black box.

In this work we introduce Viash, a tool for speeding up pipeline
prototyping through code generation, component modularity and separation
of concerns. With Viash, a user can create a pipeline module by writing
a small script or using a pre-existing code block, adding a small amount
of metadata, and using Viash to generate the boilerplate code needed to
turn it into a modular NextFlow component. This separates the component
functionality from the pipeline workflow, thereby allowing a component
developer to focus on implementing the required functionality using the
domain-specific toolkit at hand while being completely agnostic to the
chosen pipeline framework. Similarly, a pipeline developer can design a
pipeline by chaining together Viash modules while being completely
agnostic to the scripting language used in the component.

\hypertarget{core-features-and-functionality}{%
\section{Core features and
functionality}\label{core-features-and-functionality}}

\begin{figure}[t]
    \centering
    \includegraphics[width=\linewidth]{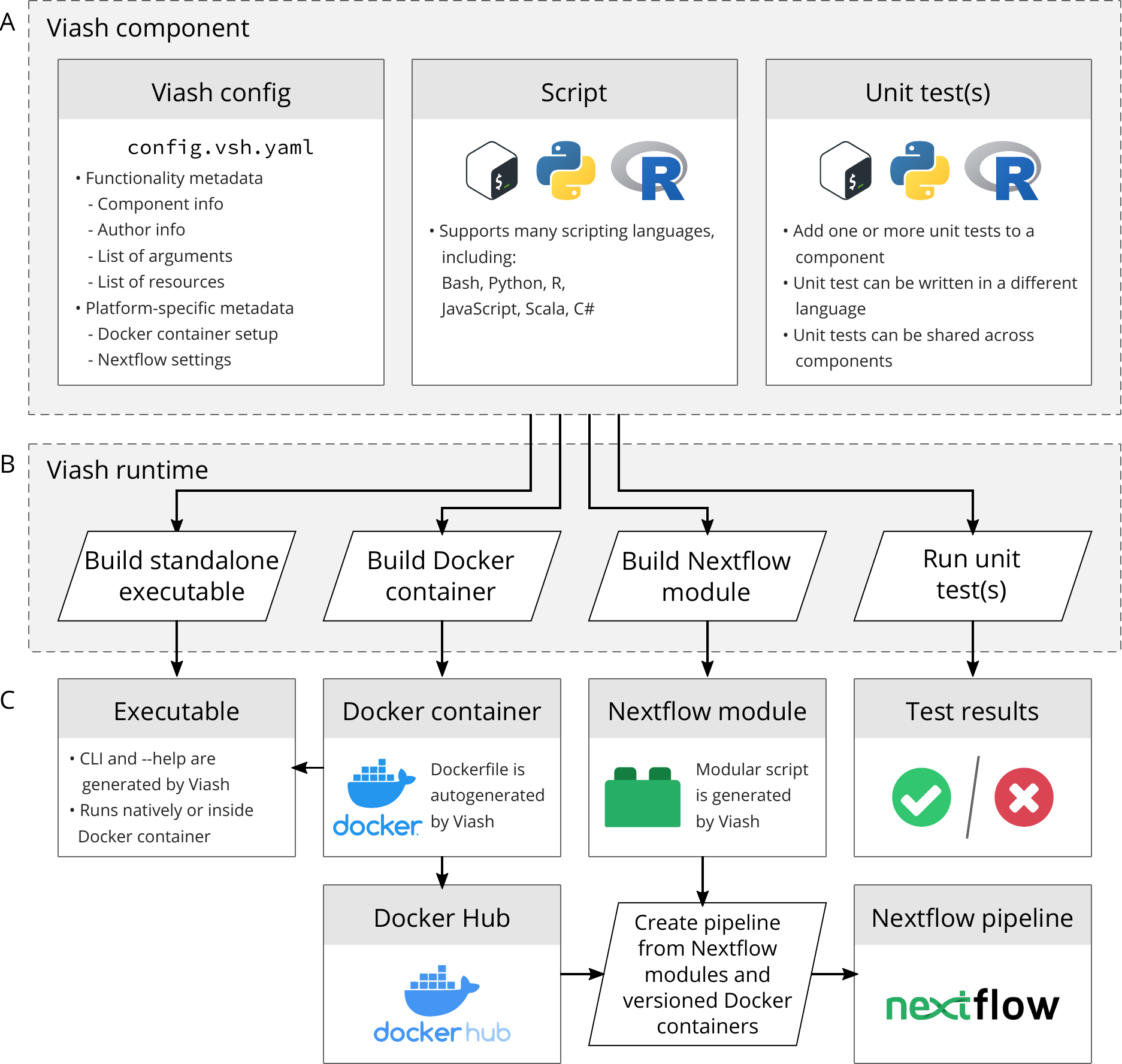}
    \caption{
      Viash allows easy prototyping of reusable pipeline components. A: Viash requires two main inputs, a script (or code block) and a Viash config file. A Viash config file is a YAML file with metadata describing the functionality provided by the component (e.g. a name and description of the component and its parameters), and platform-specific metadata (e.g. the base Docker container to use, which software packages are required by the component). Optionally, the quality of the component can be improved by defining one or more unit tests with which the component functionality can be tested. B: Viash supports robust pipeline development by allowing users to build their component as a standalone executable (with auto-generated CLI), build a Docker container to run the script inside, or turn the component into a standalone Nextflow module. If unit tests were defined, Viash can also run all of the unit tests and provide users with a report. C: By creating a versioned release (e.g. on GitHub and Docker hub), the Nextflow modules produced by Viash can be combined to create a reproducible and modular Nextflow pipeline.
    }
    \label{fig:overview}
\end{figure}

Viash is an open-source embodiment of a `code-first' concept for
pipeline development. Many bioinformatics research projects (and other
software development projects) start with prototyping functionality in
small scripts or notebooks in order to then migrate the functionality to
software packages or pipeline frameworks. By adding some metadata to a
code block or script (Figure~\ref{fig:overview}A), Viash can turn a
(small) code block into a highly malleable object. By encapsulating core
functionality in modular building blocks, a Viash component can be used
in a myriad of ways (Figure~\ref{fig:overview}B-C): export it as a
standalone command-line tool; create a highly intuitive and modular
Nextflow component; ensure reproducibility by building, pulling, or
pushing Docker containers; or running one or more unit tests to verify
that the component works as expected. Integration with CI tools such as
GitHub Actions, Jenkins or Travis CI allows for automation of unit
testing, rolling releases and versioned releases.

\begin{figure}[t]
    \centering
    \includegraphics[width=\linewidth]{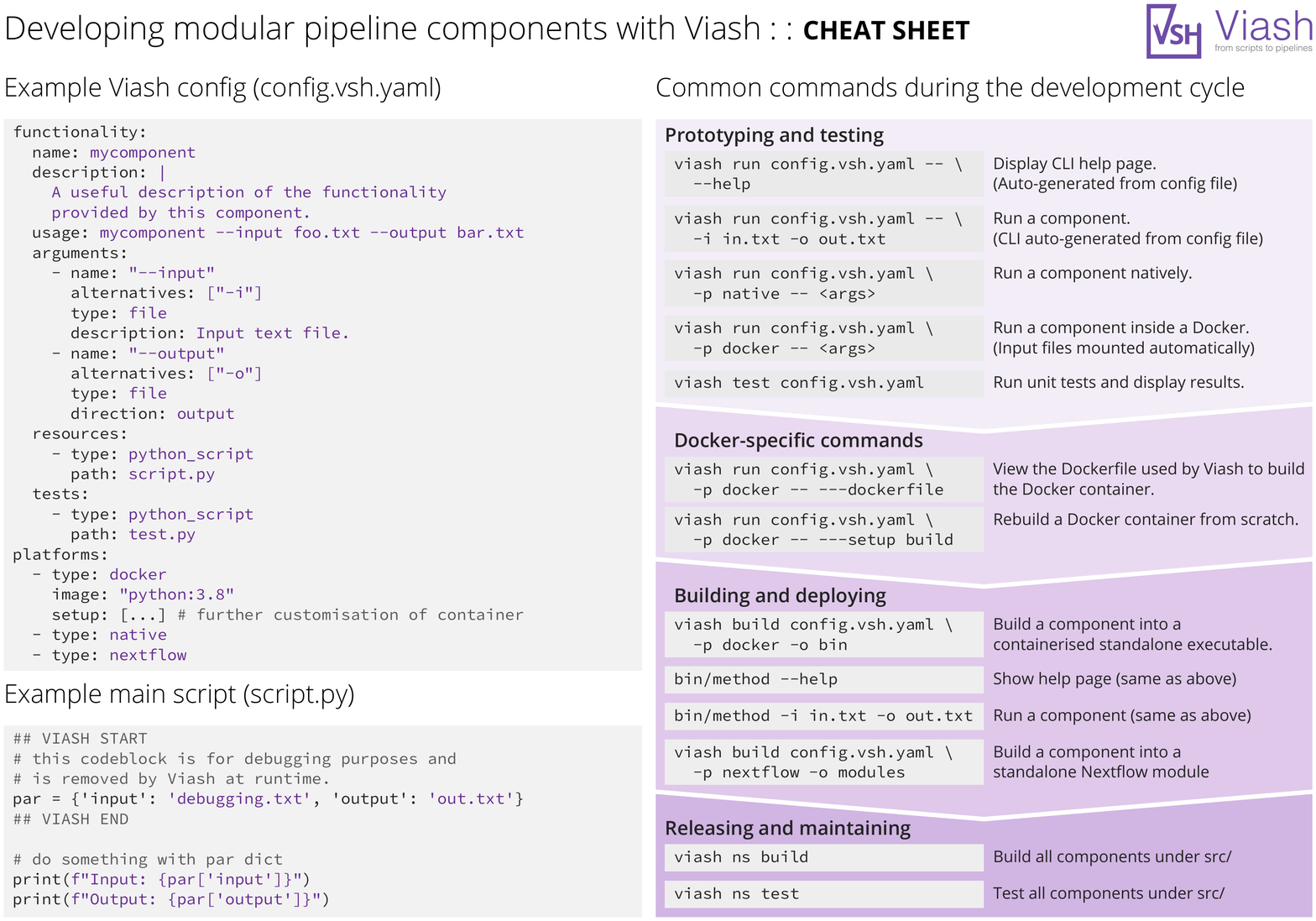}
    \caption{
      Cheat sheet for developing modular pipeline components with Viash, including a sample Viash component (left) and common commands used throughout the various stages of a development cycle (right).
    }
    \label{fig:cheatsheet}
\end{figure}

The definition of a Viash component -- a config and a code block -- can
be implemented quite concisely (Figure~\ref{fig:cheatsheet}~left). Viash
currently supports different scripting languages, including Bash,
JavaScript, Python and R. Through the use of several subcommands
(Figure~\ref{fig:cheatsheet}~right), Viash can build the component into
a standalone script using one of three backend platforms -- native,
Docker, or Nextflow. Additional commands allow processing one or more
Viash components simultaneously, e.g.~for executing a unit test suite or
(re-)building component-specific Docker images.

One major benefit of using code regeneration is that best practices in
pipeline development can automatically be applied, whereas otherwise
this would be left up to the developer to develop and maintain. For
instance, all standalone executables, Nextflow modules and Docker images
are automatically versioned. When parsing command-line arguments,
checking for the availability of required parameters, the existence of
required input files, or the type-checking of command-line arguments is
also automated. Another example is helper functions for installing
software through tools such as apt, apk, yum, pip or R devtools, as
these sometimes require additional pre-install commands to update
package registries or post-install commands to clean up the installation
cache to reduce image size of the resulting image. Here, Viash could be
the technical basis for a community of people committed to sharing
components that everybody can benefit from.

\hypertarget{applications-in-bioinformatics}{%
\section{Applications in
bioinformatics}\label{applications-in-bioinformatics}}

Ultimately, Viash aims to support pan-organisational and
interdisciplinary research projects by simplifying collaborative
development and maintenance of (complex) pipelines. While Viash is
generally applicable to any field where scalable and reproducible data
processing pipelines are needed, one field where it is particularly
useful is in single-cell bioinformatics since it supports most of the
commonly used technologies in this field, namely Bash, Python, R,
Docker, and Nextflow.

\begin{figure}[t]
    \centering
    \includegraphics[width=\linewidth]{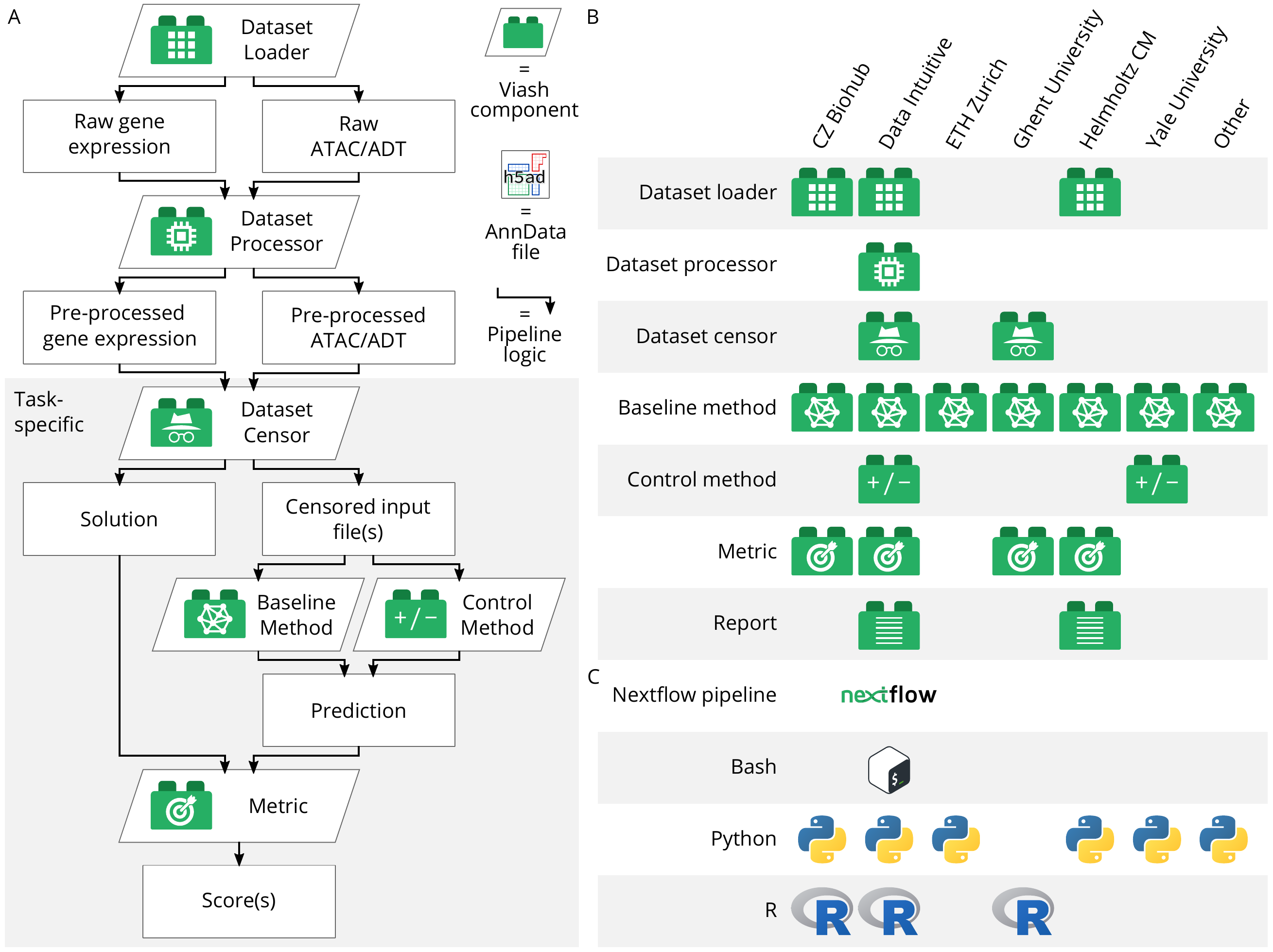}
    \caption{
      A recent NeurIPS competition for multimodal data integration\cite{luecken_sandboxpredictionintegration_2021} demonstrates the practical value of Viash by using Bash, R, Python, Docker, Nextflow, Viash, and Amazon EC2 as core technologies to run a pilot benchmark. A: The pilot benchmark pipeline consists of several types of components, each of which had strict predefined input-output interfaces. Interpreted from Luecken et al. \cite{luecken_sandboxpredictionintegration_2021}. B: Comparing which organisations contributed one or more Viash components to the pipeline demonstrates that Viash allows multiple organisations to participate in developing a pipeline collaboratively. \textbf{Note:} this visualisation pertains to one aspect of organising the NeurIPS competition, and does not at all reflect the overall efforts made by any party. C: Developers are encouraged to implement components in their preferred scripting language. Thanks to the modularity provided by Viash, sewing together multiple components into a Nextflow pipeline can be left up to a few developers, without requiring all collaborators to have expert knowledge regarding infrastructure-specific technologies.
    }
    \label{fig:usecase}
\end{figure}

A recent NeurIPS competition organised by OpenProblems
\autocite{luecken_sandboxpredictionintegration_2021} demonstrates the
practical value of Viash. As part of the preparation for the
competition, a pilot benchmark was implemented to evaluate and compare
the performance of a few baseline methods (Figure~\ref{fig:usecase}A).
By pre-defining the input-output interfaces of several types of
components (e.g.~dataset loaders, baseline methods, control methods,
metrics), developers from different organisations across the globe could
easily contribute Viash components to the pipeline
(Figure~\ref{fig:usecase}B). Since Viash automatically generates Docker
containers and Nextflow pipelines from the meta-data provided by
component developers, developers could contribute components whilst
making use of their programming environment of choice without needing to
have any expert knowledge of Nextflow or Amazon EC2
(Figure~\ref{fig:usecase}C). Thanks to the modularity of Viash
components, the same components used in running a pilot benchmark are
also used by the evaluation worker of the competition website itself. As
such, the pilot benchmark also serves as an integration test of the
evaluation worker.

\hypertarget{discussion}{%
\section{Discussion}\label{discussion}}

Viash is under active development. Active areas of development include
expanded compatibility between Viash and other technologies
(i.e.~additional scripting languages, containerisation frameworks and
pipeline frameworks), and ease-of-use functionality for developing and
managing large catalogues of Viash components (e.g.~simplified
continuous integration, allowing project-wide settings, automating
versioned releases).

We appreciate and encourage contributions to or extensions of Viash. All
source code is available under a GPL-3 license on Github at
\href{https://github.com/viash-io/viash}{github.com/viash-io/viash}.
Extensive user documentation is available at
\href{https://viash.io}{viash.io}. Requests for support or expanded
functionality can be addressed to the corresponding authors.

\printbibliography

\end{document}